\documentclass[11pt,prl,showpacs,nofootinbib,a4paper]{revtex4}
\usepackage{amsmath}
\usepackage{amsfonts}
\usepackage{amssymb}
\usepackage{graphicx} 
\usepackage{times}
\begin{document}
%
%
\date{October 3rd 2006} 
\title{THE EMERGENCE OF THE CABIBBO ANGLE \\
IN NON-DEGENERATE COUPLED SYSTEMS OF FERMIONS}
\author{Q. Duret}
\email[E-mail: ]{duret@lpthe.jussieu.fr}
\author{B. Machet}
\email[E-mail: ]{machet@lpthe.jussieu.fr}
\affiliation{Laboratoire de Physique Th\'eorique et Hautes \'Energies,
(Paris, France)}
\altaffiliation[UMR 7589 (CNRS / Universit\'e Pierre
et Marie Curie-Paris6 / Universit\'e
Denis Diderot-Paris7).\\ Postal address: ]{LPTHE tour 24-25, 5\raise 3pt
\hbox{\tiny \`eme} \'etage,
          Universit\'e P. et M. Curie, BP 126, 4 place Jussieu,
          F-75252 Paris Cedex 05 (France)} 
\begin{abstract}
Investigating, in direct continuation of our previous paper
\cite{DuretMachet1}, the implications of the 
non-unitarity of mixing matrices for non-degenerate
coupled systems that we demonstrated there, we examine more accurately
the vicinity  of Cabibbo-like mixing in quantum field theory.
We show that it is possible to preserve one of its main features, namely that,
in the space of mass eigenstates, the two  requirements -- of universality
for weak diagonal currents and --  of the absence of their non-diagonal
counterparts, although not fulfilled separately any more, can however 
reduce to a single condition for a unique mixing angle
$\theta_c$. This leads to $\tan(2\theta_c)=\pm1/2$,
or $\cos\theta_c \approx 0.9732$, only $7/10000$ away from
experimental results.  No mass ratio appears in the argumentation.
\end{abstract}
\maketitle
%

\section{Introduction}

Why mixing angles are what they are is, in addition to their more and more
precise experimental determination, one of the most active present
domain of research, both in the leptonic and hadronic sectors.
The attempts that have been 
proposed up to now in order to address this fundamental question, 
such as the linking of the sine of the Cabibbo angle \cite{Cabibbo}
 to the $d$ and $s$ quark mass ratio \cite{Weinberg}, 
are often based on empirical evidence, and so remain unsatisfying
\footnote{The rigorous treatment \cite{MachetPetcov} shows that only
weaker ``asymptotic'' relations hold, which involve both $m_d/m_s$ and
$m_u/m_c$.}
. 
Indeed, there are actually too many free parameters in the most
 general Yukawa couplings of the standard model to infer mixing angles 
from mass ratios.

In \cite{DuretMachet1}, we have demonstrated the important property that, in
quantum field theory, unlike in the Wigner-Weisskopf approximation of
quantum mechanics, the
mixing matrix of non-degenerate coupled systems of  particles 
should never be parametrized as unitary ; it
 had already been checked for neutral kaons \cite{MaNoVi}, and 
holds in particular for fermions which are, in the 
standard model, coupled to each other through the Higgs boson.
From this fondamental - although often ignored - feature, we have shown that 
two sets of solutions arise, which satisfy the combined requirements
of universality for the diagonal neutral currents of {\em mass eigenstates}
(which we will refer to hereafter as condition (C1))
 and of the absence of their off-diagonal counterparts, that we call
hereafter ``mass changing neutral currents'' -- MCNC's --  (condition (C2))
 :\newline
$\ast$\ two-parameter solutions, for which (C1) and (C2) are
independent ; two different mixing angles $\theta_1$ and $\theta_2$
 then occur ; they include in particular the so-called ``maximal mixing''
;\newline
$\ast$\ one-parameter (also called Cabibbo-like) solutions, for which (C1) and
(C2) coincide, and the two mixing angles are related by $\theta_2 =
\pm \theta_1 + k\pi$.

We shall focus here our attention on the latter,
 investigating more thoroughly their close neighborhood through 
the study of neutral weak currents of mass eigenstates at 
\begin{equation}
\theta_2 = \pm\theta_1 + \epsilon.
\label{eq:neib}
\end{equation}
While in the Cabibbo case $\theta_2 = \pm \theta_1$, (C1) is exactly 
satisfied 
together with (C2) (which is expressed, as will be shown below, 
by the vanishing of two algebraic expressions $R=0$, $S\ \text{or}\ T =0$), 
this is no more the case when one moves at (\ref{eq:neib}). 
Nevertheless, it 
is still possible to maintain the alignment of (C1) and (C2), that is 
their being approximately fulfilled with the same accuracy (which is expressed 
by the condition $|R|=|S|$ or $|R|=|T|$). It turns out that this requirement, 
which is trivial in the exact Cabbibo-case, provides  in its
 ${\cal O}(\epsilon)$ neighborhood a very accurate estimate of the
Cabbibo angle.

\section{Generalities ; states and (weak) currents}

In quantum field theory, the states of definite mass for a system of particles 
are defined as the eigenstates, at its poles, of the full (renormalized) 
propagator, which is given by a matrix of dimension $2n_f$ in flavour space.
For non-degenerate coupled systems, like neutral kaons, leptons or quarks,
these mass eigenstates belong to different bases \cite{DuretMachet1}.
Indeed, one finds only one such eigenstate at each of the poles of 
the propagator ; it is, among the $2n_f$ eigenstates of the latter, 
the one corresponding to the vanishing eigenvalue. Hence, the mass eigenstates
 do not make up an orthonormal basis, and
the mixing matrix, which by definition connects the
set of mass eigenstates to that of flavour eigenstates, is non-unitary. 

Nevertheless, for systems of coupled fields which, like quarks,
are never on-shell,  a natural basis appears : the one that occurs at
 any given $z=q^2$. It is then
orthonormal as soon as the inverse propagator $L^{(2)}(z)$ is hermitian.
Thus, the mixing matrix relating it to the flavour basis is unitary 
(assuming that the flavour basis is orthonormal, too). 
This has, in the simple case of two generations, 
the following consequences.
While, in general, ($2 \times 2$) mixing matrices for coupled systems 
are to be parametrized with two mixing angles $\theta_1$ and $\theta_2$,
quark-like (Cabibbo-like) systems finally shrink to one dimension.

In the following we shall mainly deal with weak currents in the space of 
mass eigenstates which, as stressed in \cite{DuretMachet1}, underlie
the physics of mixing angles.
Neutral (left-handed) weak currents for mass eigenstates are determined by the
combinations $K_1^\dagger K_1$ and $K_2^\dagger K_2$, where $K_1$ and $K_2$
are the mixing matrices for the two types of fermions concerned
(for example neutral and charged leptons or 
quarks of the $u$ and $d$-type), whereas charged currents involve
the combination $K=K_1^\dagger K_2$.
As soon as neither $K_1$ nor $K_2$ is unitary, the conditions for
universality  (equality of diagonal neutral currents) and for the
 absence of off-diagonal neutral currents, which are built-in properties
in flavour space, appear no longer trivial in the space
of mass eigenstates, but give rise to specific constraints which 
we will study in detail.

We will only deal hereafter with $K_1$, but the process would 
be exactly the same for $K_2$.\\
Let us parametrize $K_1$ like in \cite{DuretMachet1}, with two mixing angles
$\theta_1$ and $\theta_2$ :
\begin{equation}
K_1 = \left(\begin{array}{rr} e^{i\alpha}c_1 & e^{i\delta}s_1 \cr
-e^{i\beta}s_2 & e^{i\gamma}c_2
\end{array}\right).
\label{eq:Kparam}
\end{equation}
In the mass basis, the weak neutral current couplings of the
weak Lagrangian write
\begin{equation}
\overline{\Psi}_m \gamma^\mu \Big(\frac{1-\gamma^5}{2} \Big)
W_\mu^3 K_1^\dagger K_1 \Psi_m,
\label{eq:NWC}
\end{equation}
where $\Psi$ may stand for 
$\left(\begin{array}{c} u_m \cr c_m \end{array}\right)$.
From the explicit expression of $K_1^\dagger K_1$ obtained with
(\ref{eq:Kparam}), it is straightforward to see that
(C1) imposes the vanishing of
\begin{equation}
R = c_1^2 + s_2^2 - c_2^2 -s_1^2,
\label{eq:univ}
\end{equation}
while (C2) requires that of
\begin{equation}
S = c_1s_1 -c_2s_2\quad \text{or}\quad T = c_1s_1 + c_2s_2.
\label{eq:mcnc}
\end{equation}
%

\section{The emergence of the Cabibbo angle}

Due to the smallness of the expected deviation of $K_1$ from unitarity,
we shall only explore the vicinity (\ref{eq:neib}) of  the Cabibbo solutions
\footnote{The ``$+$'' sign corresponds, there, to the condition $S$ and the ``$-$''
 sign to $T$.}
.

Calling hereafter $\theta_1 = \theta_c$,
$K_1^\dagger K_1$ expands then at ${\cal O}(\epsilon)$ as
\begin{equation}
K_1^\dagger K_1 = \left(\begin{array}{cc} 1 & 0 \cr
0 & 1
\end{array}\right) + \epsilon
\left(\begin{array}{cc} \sin(2\theta_c) & - a \cos(2\theta_c) \cr
-a^\ast \cos(2\theta_c) & -\sin(2\theta_c)
\end{array}\right)
+{\cal O}(\epsilon^2),
\label{eq:KdagK}
\end{equation}
where $a=e^{i\theta_a}$ is a complex number of unit modulus related to
the phases $\alpha, \beta, \gamma, \delta$ present
in the original parametrization (\ref{eq:Kparam}) of $K_1$
\footnote{See also section $3.2.1$ of \cite{DuretMachet1} which explains 
how these phases are linked to each other by the requirement of the
absence of MCNC's.}
.
As for $|R|$, $|S|$ and $|T|$, they respectively become
\footnote{The expressions for $S$ and $T$ can be straightforwardly obtained
from the relations called $\text{f}_1$ and $\text{g}_1$ in
\cite{DuretMachet1} for
$\theta_2$ in the vicinity of $\theta_1$; the one for $R$ is likewise
obtained in the same conditions from the relation in the line just above.}
\footnote{The identical normalization of the two diagonal terms
proportional to $\epsilon$ in (\ref{eq:KdagK}) is fortuitous, as shows the
case of three generations \cite{DuretMachet3}; universality is thus 
expressed on general grounds by the vanishing of $(2\sin\theta_c)$, 
which represents the {\em
difference} between the two diagonal terms, and not by $\zeta
\sin\theta_c$, where $\zeta$ would be an arbitrary number.}
\begin{equation}
|R| = 2\epsilon |\sin 2\theta_c|,\quad |S| = |T| = \epsilon |\cos 2\theta_c|.
\end{equation}
One obviously cannot ensure at ${\cal O}(\epsilon)$
 the simultaneous fulfillment of 
(C1) and (C2). However it is still possible to preserve one important
feature of the Cabbibo-like solutions, i.e. that 
the two conditions (C1) and (C2) reduce to a single one, which translates into
\begin{equation}
\tan(2\theta_c) = \pm \frac{1}{2}.
\label{eq:tan2}
\end{equation}
By the change of variables \cite{Pestieau}
 $y = \tan(\pi/4 -\theta_c) \Leftrightarrow
\tan\theta_c = \displaystyle\frac{1-y}{1+y}$, (\ref{eq:tan2}) becomes
equivalent to
\footnote{Equations of this type were empirically advocated for in
\cite{Pestieau1}.}
\begin{equation}
\displaystyle\frac{1}{y} - y = \pm 1,
\label{eq:gold}
\end{equation}
one of the solutions of which is the golden number
\begin{equation}
\varphi = \displaystyle\frac{1+\sqrt{5}}{2}.
\label{eq:gold1}
\end{equation}
The four solutions of (\ref{eq:tan2}) can then be rewritten 
in terms of $\varphi$ :
\begin{equation}
\tan\theta_c =  \pm \displaystyle\frac{\varphi-1}{\varphi+1},
\  \pm \displaystyle\frac{\varphi+1}{\varphi-1}.
\label{eq:gold2}
\end{equation}
The first two, which have opposite signs, correspond to
$\cos\theta_c = 0.9732$;
this lies only $7/10000$ away from the present \cite{PDG} experimental range
$[0.9739, 0.9751]$ for the Cabibbo angle generally refered to.

Eq. (\ref{eq:gold}) is symmetric by $y\to 1/y$ and $y\to -y$ (and likewise
by their combination $y \to -1/y$).
When $y \to 1/y$, $\theta_c \to -\theta_c$; when
$y \to -y$, $\theta_c \to (\pi/2 -\theta_c)$ or $\theta_c \to
-(\theta_c + \pi/2) \equiv -(\pi/2 -(-\theta_c))$; when $y \to -1/y$,
$\theta_c \to \theta_c \pm \pi/2$. 
Owing to the invariance of the $\tan$ function by a translation by $\pi$ of
its argument, the set of solutions of (\ref{eq:gold}) inside the $[0,2\pi]$
interval appears as described on Fig.~1.

\vskip 5mm

\vbox{
\begin{center}
\includegraphics[height=6truecm,width=6truecm]{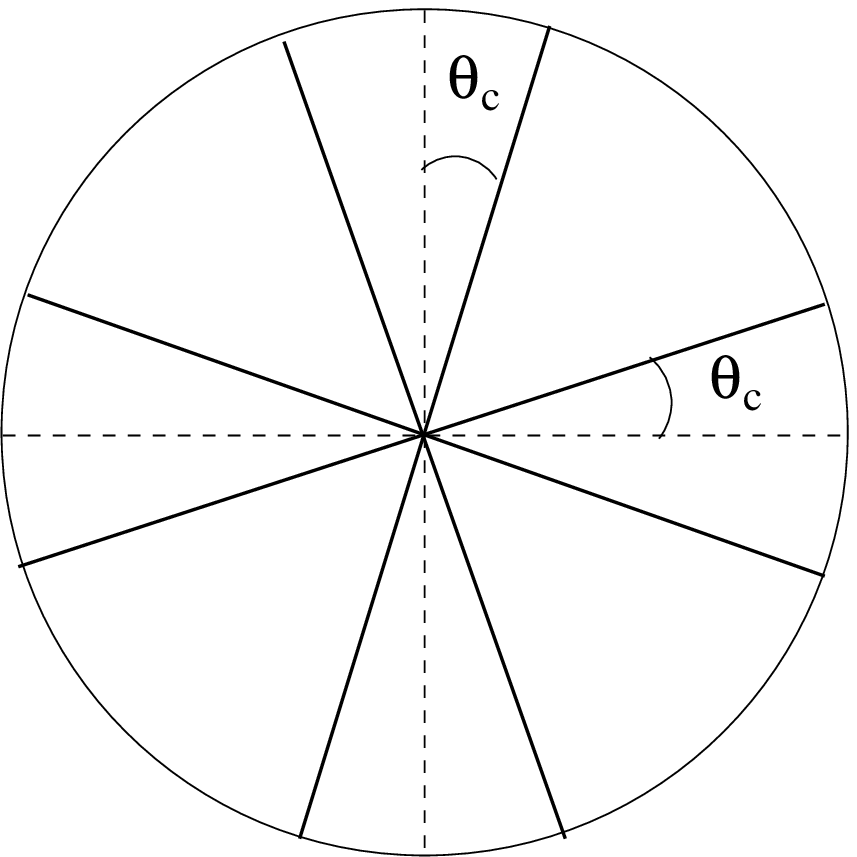}
\end{center}

\centerline{{\em Fig.~1: solutions for the Cabibbo angle}}
}

\vskip 5mm

The physical meaning of these symmetries appears on (\ref{eq:NWC}).
It is invariant by the change $\theta_c \to -\theta_c$ if the 
following (unitary) transformation of the fields is simultaneously performed :
\begin{equation}
u_m \to e^{i(\theta_a \mp \frac{\pi}{2})}c_m \qquad \mathrm{and} \qquad 
c_m \to e^{i(-\theta_a \mp \frac{\pi}{2})}u_m.
\end{equation}

Likewise, (\ref{eq:NWC}) is invariant under the change
$\theta_c \to \frac{\pi}{2}-\theta_c$  with
\begin{equation}
u_m \to e^{\pm i \frac{\pi}{2}}u_m \qquad \mathrm{and} \qquad 
c_m \to e^{\mp i \frac{\pi}{2}}c_m.
\end{equation}  
Thus, the symmetry $y \to 1/y$ of (\ref{eq:gold}) reflects the 
invariance of neutral currents of mass eigenstates by the exchange
of families $u_m \leftrightarrow c_m, d_m \leftrightarrow s_m$, and, hence,
the condition of universality for such currents, 
while the symmetry $y \to -y$ corresponds to a simple rephasing of the
fields
\footnote{The interpretation above is deduced from considering  weak
{\em currents}; if one instead considers the mixing matrix of fermionic
fields, the role of the symmetries is swapped:
 the transformation $\theta_c \to \pi/2 - \theta_c$, or $y \to -y$
gets associated with the exchange of families, while the transformation $y
\to 1/y \Leftrightarrow \theta_c \to -\theta_c$ corresponds to a simple
rephasing of the fields. Physically, it has however always been known that the
currents are the relevant quantities.}.

This approach yields a {\em constant} value for the Cabibbo
angle, a feature which should be confronted with experiment.
The most natural, {\em a
priori} $q^2$-dependent, orthonormal basis related to Cabibbo-like systems
mentioned in section 2, should then also exhibit special
properties with respect to its $q^2$-dependence. This will be investigated
in a forthcoming work.

\section{Conclusion and prospects}

Realizing, as was first done in \cite{MaNoVi}, that mixing matrices of
non-degenerate coupled systems should not be parametrized as unitary, led
in \cite{DuretMachet1} to uncover maximal mixing of leptons as
a class of solutions of very simple physical conditions for their
mass eigenstates.

We have now shown that the measured value of the Cabibbo angle $\theta_c$
is the one ensuring, in its first order vicinity, the property,
already stressed in \cite{DuretMachet1}, that
universality of diagonal neutral currents for mass eigenstates and the
absence of MCNC's reduce to a unique condition.

In this elementary algebraic calculation, no mass ratio appears.
It may thus help to provide independent information on the latter.
On another side, this feature is  welcome for quark-like systems which
cannot be defined on-shell and for which, accordingly, the notion of
physical mass is ill-defined.

This work, together with \cite{DuretMachet1}, strongly suggests that the
observed values of mixing angles for quarks and leptons follow from
simple physical requirements.
The generalization to three generations is currently under investigation
\cite{DuretMachet3}.

%
\begin{em}

\end{em}

\end{document}